\RequirePackage{ifpdf}
\documentclass[12pt,letterpaper]{JHEP3}         
\usepackage{amssymb,amsfonts}

\usepackage{wrapfig}
\usepackage{epsfig}

\def\be{\begin{eqnarray}}
\def\ee{\end{eqnarray}}
\newcommand{\nn}{\nonumber}
\newcommand\para{\paragraph{}}

\newcommand{\eqn}[1]{(\ref{#1})}

\newcommand{\vphi}{\varphi}

\def\Dslash{\,\,{\raise.15ex\hbox{/}\mkern-12mu D}}
\def\Dbarslash{\,\,{\raise.15ex\hbox{/}\mkern-12mu {\bar D}}}
\def\delslash{\,\,{\raise.15ex\hbox{/}\mkern-9mu \partial}}
\def\delbarslash{\,\,{\raise.15ex\hbox{/}\mkern-9mu {\bar\partial}}}
\def\pslash{\,\,{\raise.15ex\hbox{/}\mkern-9mu p}}
\def\calDslash{\,\,{\raise.15ex\hbox{/}\mkern-12mu {\cal D}}}

\newcommand{\ket}{\rangle}

\newcommand{\up}{|\uparrow\rangle}
\newcommand{\down}{|\downarrow\rangle}
\def\ket#1{|{#1}\rangle}

\def\lae{\mathrel{\mathop{\smash{\lower .5 ex \hbox{$\stackrel<\sim$}}}}}
\def\lae{\mathrel{\mathop{\smash{\lower .5 ex \hbox{$\stackrel>\sim$}}}}}

\title{A Matrix Model for Non-Abelian Quantum Hall States}

\author{Nick Dorey$^1$, David Tong$^{1,2}$ and Carl Turner$^1$\\
$^1$Department of Applied Mathematics and Theoretical Physics, \\
University of Cambridge, \\
Cambridge, CB3 OWA, UK \\ 
$^2$Department of Theoretical Physics \\
Tata Institute of Fundamental Research \\
Homi Bhabha Road, Mumbai 400 005, India 

{\tt  n.dorey, d.tong, c.p.turner@damtp.cam.ac.uk}\\}

\abstract{We propose a matrix quantum mechanics for  a class of non-Abelian quantum Hall states. The model describes  electrons which carry an internal  $SU(p)$ spin. The ground states of the matrix model include spin-singlet generalisations of the Moore-Read and Read-Rezayi states and, in general, lie in a class previously introduced by  Blok and Wen.  The effective action for these states is a  $U(p)$ Chern-Simons theory. We show how the matrix model can be derived from quantisation of the vortices in this Chern-Simons theory and how the matrix model ground states can be reconstructed as  correlation functions in the  boundary WZW model.}

\begin{document}
\pagestyle{plain} \setcounter{page}{1}
\newcounter{bean}
\baselineskip16pt \setcounter{section}{0}

\section{Introduction and Summary}


Electrons in the lowest Landau level exhibit an astonishing array of compressible and incompressible states, the latter with both Abelian and non-Abelian topological order. For a number of these states, a description of the dynamics in terms of {\it matrix models} has proven useful. These typically involve a $U(N)$ gauge symmetry, with $N$ the number of electrons. The gauge symmetry imposes constraints on the Hilbert space which project the dynamics onto the lowest Landau level.

\para
The first such matrix model was developed by Pasquier and Haldane \cite{pasquier} (see also \cite{readmatrix}) to describe the compressible state at half-filling. Subsequently, Polychronakos introduced a matrix model for the Laughlin states \cite{alexios}, inspired by earlier work \cite{susskind}. The Hilbert space of this matrix model not only lies in the lowest Landau level, but also captures the appropriate topological order. The purpose of this paper is to introduce a generalisation of these matrix models describing a class of non-Abelian quantum Hall states.


\subsection*{A Class of Non-Abelian Quantum Hall States}

Before we describe the role played by the matrix model, we first summarise some properties of the non-Abelian Hall states that will emerge. 

\para
The original Moore-Read state \cite{mr}, and its extension to the series of Read-Rezayi states \cite{rr}, describe spin polarised electrons. There are, however, a number of prominent non-Abelian Hall states in which the electrons carry an internal spin degree of freedom \cite{bw,ardschout,as2}. Typically, the quantum Hall ground states are singlets under the spin symmetry group. It is this kind of ``non-Abelian spin-singlet state" which will be of interest in this paper. 

\para
In the context of quantum Hall physics, the ``spin" degrees of freedom  can be more general than the elementary spin of the electron. For example, in bilayer systems the layer index plays a similar role to the  spin degree of freedom and is sometimes referred to as a ``pseudospin". In other systems, the electrons may carry more than two internal states. This occurs, for example, in graphene where one should include both spin and valley degrees of freedom \cite{dean}. Here, we will consider systems in which each particle carries some  number of internal states.  This will include situations in which these states transform in a higher representation of $SU(2)$, but also situations  in which the states transform under a general $SU(p)$ group. In all cases, we will refer to these internal states simply as the ``spin" degrees of freedom of the particle.

\para
When the symmetry group is  $SU(2)$, one can construct non-Abelian spin-singlet states starting from the familiar Abelian  $(m,m,n)$ Halperin states \cite{halperin}. It is well known that when particles carry spin $s=\frac{1}{2}$, only the Halperin states with $m=n+1$ are spin-singlets \cite{haldane}. Apparently less well-known is the statement that for particles carrying spin $s$, the $(m,m,n)$ states, suitably interpreted,  are spin-singlets when $m=n+2s$.  Moreover, the presence of the spin degrees of freedom changes the universality class of these states and, for $s> \frac{1}{2}$, they have non-Abelian topological order. In particular, when the particles have spin $s=1$, it is possible to rewrite these states in Pfaffian form and they lie in the same universality class as the Moore-Read states.


\para
When the symmetry group is $SU(p)$, the obvious $(m,\ldots,m,n\ldots,n)$ generalisation of the Halperin states can again be used as the foundation to build non-Abelian states. When $m-n=1$, these are spin-singlets if each particle transforms in the fundamental representation of $SU(p)$. More generally, when $m-n=k$ one can build spin-singlets if each particle transforms in the $k^{\rm th}$ symmetric representation of $SU(p)$. 

\para
The states that arise in this way are not novel. They were first introduced many  years ago by Blok and Wen \cite{bw}, albeit using the rather different construction of conformal blocks in an  $SU(p)_k$ WZW model. The states have filling fraction
\be \nu = \frac{p}{k+pn}\label{thatnow}\ee
with $p$ and $k$ positive integers determined by the spin group and its representation, and $n$ an arbitrary positive integer.
For $p=1$, these  are simply the Laughlin states. For $p=k=2$, these are spin-singlet generalisations of the Moore-Read states. For $p>2$ and $k=2$, these are spin-singlet generalisations of the Read-Rezayi states. 


\subsection*{Chern-Simons Theories and  Matrix Models}

The effective description of the Blok-Wen states is a non-Abelian Chern-Simons theory. The gauge group and levels are given by
\be U(p)_{(k+pn)p,k}= \frac{U(1)_{(k+pn)p} \times SU(p)_k}{{\bf Z}_p}\label{thisnow}\ee
The allowed level of the $U(1)$ factor is strongly constrained by the fact that this is a $U(p)$ rather than $U(1)\times SU(p)$ theory  \cite{sw}.

\para
Viewed in a certain slant of light,  the Blok-Wen states are the most natural non-Abelian quantum Hall states. Let us take a quick aside to explain this.  The long-distance physics of all  non-Abelian quantum Hall states is described by some variant of non-Abelian Chern-Simons theories. This means, of course, that Wilson lines in this theory carry some representation under the non-Abelian group which, for us, is $SU(p)$. The corresponding ``colour" degrees of freedom are then interpreted as spin degrees of freedom of the underlying electron. This, in essence, is why non-Abelian quantum Hall states arise naturally from particles carrying internal spin.

\para
In contrast, if one wants to describe the long-distance physics of spin-polarised non-Abelian Hall states, such as those of  \cite{mr,rr}, one must work somewhat harder. This involves the  introduction of yet further quotients of the 3d Chern-Simons theory \cite{sw} to eliminate the spin degrees of freedom. This is the sense in which the Blok-Wen states  are particularly natural\footnote{Things look somewhat different when viewed from the boundary perspective. The same quotient that appears complicated in the 3d bulk can result in a very simple boundary theory, such as the Ising  \cite{mr} or parafermion \cite{rr} conformal field theories. }.

\para
We are now in a position to explain how the matrix model arises. The electrons in the quantum Hall system correspond to  vortices of the $U(p)$ Chern-Simons theory. The $U(N)$ matrix model is simply the description of the microscopic dynamics of $N$ of these vortices. At present, we can construct this matrix model only for the choice $n=1$ in \eqn{thatnow} and  \eqn{thisnow}. It is to be expected that quantising these vortices results in the quantum Hall ground state. The matrix model provides the technology to do this explicitly.

\para
This relationship between Chern-Simons vortices, matrix models and quantum Hall wavefunctions was explored for Abelian theories in  \cite{unknown,us}. The novelty in non-Abelian gauge theories is that the vortices are endowed with an internal orientation, or spin, degrees of freedom, as first explained in \cite{me,yung}.  We will show that this results in the non-Abelian quantum Hall states described above.  (An earlier, somewhat orthogonal attempt to describe a quantum Hall fluid of non-Abelian vortices was made in \cite{kimura}.)

\subsection*{Plan of the Paper}

The paper is written in a somewhat different order from the preceding introduction. In Section \ref{matrixsec}, we introduce the matrix model but do not explain its Chern-Simons origins. Instead, we will take the matrix model as the starting point and show that it describes particles with spin moving in the lowest Landau level. We will see that, upon quantisation, the ground state lies in the same universality  class as the non-Abelian quantum Hall states previously introduced by Blok and Wen \cite{bw}.

\para
Section \ref{bwsec} can be read independently. We describe in some detail the Blok-Wen wavefunctions and their construction from spin-generalisations of the Halperin-type states. We show, in particular, how they describe spin-singlet generalisations of the Moore-Read \cite{mr} and Read-Rezayi \cite{rr} states. One rather cute fact is that the Read-Rezayi states arise in this picture from  $SU(p)_2$ Chern-Simons theory; this is related by level-rank duality to the more familiar $SU(2)_p$ coset constructions.

\para
In Section \ref{vortexsec}, we return to the origin of the matrix model. We explain how it captures the dynamics of vortices in a Chern-Simons theory with gauge group \eqn{thisnow}.

\para
Finally, in Section 5, we complete the circle of ideas. We confirm  that the matrix model wavefunctions, derived from the Chern-Simons theory, can be reconstructed as correlation functions in the boundary WZW model with algebra \eqn{thisnow}.  In a follow-up paper \cite{later}, we will make the connection between the matrix model and the WZW model more direct: we will show that their partition functions agree.

\section{The Quantum Hall Matrix Model}\label{matrixsec}

The purpose of this paper is to study a matrix model description of non-Abelian  quantum Hall states.  The model will describe $N$ particles which we refer to as ``electrons". 

\para
The matrix model is a $U(N)$ gauged quantum mechanics, with a gauge field which we denote as $\alpha$. This gauge field is coupled to an $N\times N$ complex matrix $Z$, together with a set of $N$-dimensional vectors $\varphi_i$ which are labelled by an index $i=1,\ldots,p$. These transform under the gauge symmetry as 
\be Z\rightarrow  UZU^\dagger\ \ \ {\rm and}\ \ \ \varphi_i\rightarrow U \varphi_i \ \ \ \mbox{for} \  \ \ U\in U(N)\label{transform}\ee
The dynamics  is governed by the first-order action
\be S  = \int dt\ \frac{iB}{2}\,{\rm Tr}\,\left(Z^\dagger {\cal D}_t Z\right) + i\sum_{i=1}^p \varphi_i^\dagger {\cal D}_t\varphi_i - (k+p)\,{\rm Tr}\,\alpha- {\omega}\, {\rm Tr}\,Z^\dagger Z \label{matrix}\ee
with ${\cal D}_t Z = \partial_t{Z} - i[\alpha,Z]$ and ${\cal D}_t\varphi_i = \partial_t{\varphi}_i - i \alpha\varphi_i$. 

\para
The action depends on three parameters: $B$, $\omega$ and $k$. We will see below that $B$ is interpreted as the background magnetic field in which the electrons move, while $\omega$ is the strength of a harmonic trap which encourages the electrons to cluster close to the origin. Finally $k$, which  appears in the combination $k+p$, is the coefficient of the quantum mechanical Chern-Simons term. Gauge invariance requires that $k$ is an integer and we will further take it to be positive: $k\in {\bf Z}^+$.

\para
In addition to the $U(N)$ gauge symmetry, our model also enjoys an $SU(p)$ global symmetry, under which the $\varphi_i$ rotate.
When $p=1$, this action reduces to the model written by Polychronakos \cite{alexios} to describe Laughlin states. The $p=1$ matrix model was further studied in a number of papers, including \cite{hellvram,ks,ks2} and we  will review some of its properties below. The model with general $p$ was previously discussed in \cite{polywilson}, albeit with a different interpretation from that offered here.

%

\subsubsection*{Getting a Feel for the Matrix Model}

To gain some intuition for the physics underlying \eqn{matrix}, let's first look at the example of a single particle, so $N=1$ and our matrix model is an Abelian $U(1)$ gauge theory, with dynamics
\be S_{N=1} = \int dt\ \frac{iB}{2}Z^\dagger \dot{Z} + \sum_{i=1}^pi\varphi_i^\dagger{\cal D}_t\varphi_i - (k+p) \alpha - \omega\, Z^\dagger Z\nn\ee
In this case, the $Z$ field decouples; the kinetic term, which is first order in time,  describes the low-energy dynamics of an electron moving in a large external magnetic field $B$. When we come to the quantum theory, this will translate into the statement that the  electron lies in the lowest Landau level. The term proportional to $\omega$ provides a harmonic trap for the electron.

\para
Meanwhile, the $\varphi_i$ variables describe the internal degrees of freedom of the electron. To see this, note that the equation of motion for $\alpha$ requires that  $\sum_i |\varphi_i|^2 = k+p$ is constant. After dividing out by $U(1)$ gauge transformations, $\varphi_i\rightarrow e^{i\theta}\varphi_i$, we see that $\varphi_i$ parameterise the space ${\bf CP}^{p-1}$. However, the action is first order in time derivatives, which means that ${\bf CP}^{p-1}$  should be viewed as the {\it phase space} of the system, as opposed to the configuration space. This is important. Because the phase space has finite volume, the  quantisation of $\varphi_i$ will result in a finite-dimensional internal Hilbert space for the electron. In other words, the electron carries ``spin".

\para
Note that our usage of the word ``spin" is somewhat more general than its standard meaning in condensed matter physics (or high energy physics for that matter). Usually, one thinks of spin as referring to a representation of $SU(2)$; this corresponds to the choice $p=2$ in our model. More generally, our internal degree of freedom transforms in some representation of $SU(p)$. The choice of  representation is determined by the parameter $k$. (We will show below that the electrons sit in the $k^{\rm th}$ symmetric representation of $SU(p)$; in the case of $SU(2)$, this means that they carry spin $j=k/2$.)

\para
We learn that the $U(1)$ matrix model describes a particle carrying spin, restricted to move in the lowest Landau level. The $U(N)$ matrix model simply describes $N$ such particles. Roughly speaking, the $N$ eigenvalues of the matrix $Z$ correspond to the positions of the particles although, as we will see, there is some ambiguity in this when the particles are close. More precisely, we can again look at the equation of motion for the gauge field $\alpha$. This results in the $u(N)$-valued constraint
\be \frac{B}{2}[Z,Z^\dagger] +  \sum_{i=1}^p\vphi_i\vphi_i^\dagger = (k+p){\bf 1}_N\label{constraint}\ee
The phase space, ${\cal M}$, of the theory is now the space of  solutions to \eqn{constraint}, modulo the gauge action \eqn{transform}. This has real dimension ${\rm dim}\,{\cal M} = 2Np$. Our task is to quantise this phase space, with the harmonic potential $H=\omega \, {\rm Tr}\,Z^\dagger Z$ providing the Hamiltonian.

\subsection{Quantisation}\label{quantsec}

In this section, we study the quantisation of our matrix model \eqn{matrix}.  The canonical commutation relations inherited from the action \eqn{matrix} are
\be \frac{B}{2}[Z_{ab},Z^\dagger_{cd} ] = \delta_{ad}\delta_{bc}\ \ \ {\rm and}\ \ \ [\varphi_{i\,a},\varphi_{j\,b}^\dagger]= \delta_{ab}\delta_{ij}\label{commute}\ee
with $a,b=1,\ldots,N$ and $i,j=1,\ldots, p$. 
We choose a reference state $|0\rangle$ obeying
\be Z_{ab}|0\rangle = \varphi_i |0\rangle=0\nn\ee
The Hilbert space is then constructed in the usual manner by acting on $|0\rangle$ with $Z^\dagger$ and $\varphi_i^\dagger$.

\para
However, we still need to take into account the $U(N)$ gauge symmetry. This is implemented by requiring that all physical states obey the quantum version of the Gauss' law constraint \eqn{constraint}. Normal ordering the terms in the matrix commutator, this reads
\be \frac{B}{2}:[Z, Z^\dagger]: + \sum_{i=1}^p\varphi_i\varphi_i^\dagger  = (k+p){\bf 1}_N\label{qgauss}\ee
The traceless part of this equation is interpreted as the requirement that physical states are $SU(N)$ singlets. Meanwhile, the trace of this constraint requires all physical states to carry fixed charge under $U(1)\subset U(N)$. Here there is an ordering issue. Using the commutation relations \eqn{commute}, we find
\be \sum_{a=1}^N \sum_{i=1}^p\varphi_{i\,a} \varphi^\dagger_{i\,a} = (k+p)N \ \ \ \Rightarrow\ \ \ \sum_{a=1}^N \sum_{i=1}^p\varphi^\dagger_{i\,a} \varphi_{i\,a} = k N\label{traceconstraint}\ee
This tells us that all physical states carry charge $kN$ under the $U(1)$. In other words, all states in the physical Hilbert space contain  precisely $kN$ copies of $\varphi^\dagger$ acting on $|0\rangle$.

\subsection*{The Spin of the Particle Revisited}

We can now be more precise about the internal $SU(p)$ spin carried by each particle.  Setting $N=1$, the spin states of a single particle take the form
\be |\Omega_{i_1\ldots i_k}\rangle = \varphi_{i_1}^\dagger\ldots \varphi_{i_k}^\dagger|0\rangle\nn\ee
Since each operator $\varphi_i$ transforms in the fundamental of $SU(p)$, the spin states $|\Omega\rangle$ transform in the $k^{\rm th}$ symmetric representation. In particular, for $k=1$ the electrons carry the fundamental representation of $SU(p)$.

\para
Our main focus in this paper will be on quantum Hall states which are $SU(p)$ spin-singlets. Some simple group theory tells us that for this to happen we must have the number of electrons $N$ divisible by $p$. Indeed, we will see below that the ground states simplify in this case.

\subsection{The Ground States}\label{groundsec}

The ground state of the matrix model with $p=1$ was constructed in \cite{alexios}. We first review this example before explaining the straightforward generalisation to $p>1$.

\subsubsection*{The $p=1$ Ground State}

When $p=1$, the electrons carry no internal spin. The constraint \eqn{traceconstraint} tells us that all physical states have $kN$ operators $\varphi^\dagger$ acting on $|0\rangle$. Further, the Hamiltonian arising from \eqn{matrix} is 
\be H = \omega \, {\rm Tr}\,Z^\dagger Z\label{ham}\ee
which simply counts the number of $Z^\dagger$ operators acting on $|0\rangle$. The route to constructing the ground state is then straightforward: we need to act with $kN$ copies of $\varphi^\dagger$, keeping the number of $Z^\dagger$ operators to a minimum. The subtleties arise from the requirement that the physical  states are invariant under $SU(N)$ gauge transformations. Since we only have $\varphi^\dagger$ operators to play with, the only way to achieve this is to construct a baryon operator of the form
\be \epsilon^{a_1\ldots a_N} (Z^{l_1}\varphi)_{a_1}^\dagger\ldots (Z^{l_N}\varphi)_{a_N}^\dagger\nn\ee
However, because $\varphi$ is bosonic, the antisymmetrisation inherent in $\epsilon^{a_1\ldots a_N}$ causes this operator to vanish unless all the exponents $l_a$ are distinct. Because we pay an energy cost \eqn{ham} for each insertion of $Z^\dagger$, it follows that  the lowest energy operator is given by
\be 
\epsilon^{a_1\ldots a_N} (Z^{0}\varphi)_{a_1}^\dagger (Z\varphi)^\dagger_{a_2}\ldots (Z^{N-1}\varphi)_{a_N}^\dagger\nn\ee
The trace constraint then tells us that the ground state is given by
\be |{\rm ground}\rangle_k = \left[\epsilon^{a_1\ldots a_N} (Z^{0}\varphi)_{a_1}^\dagger (Z\varphi)^\dagger_{a_2}\ldots (Z^{N-1}\varphi)_{a_N}^\dagger\right]^k|0\rangle\nn\ee
The interplay between the gauge symmetry and the Hamiltonian has resulted in the construction of a state with interesting correlations between the positions of particles, encoded in the operator $Z$. This will become increasingly apparent as we proceed; in particular, shortly we will write these states in the more familiar language of $N$-particle wavefunctions and see a close relationship to the Laughlin wavefunctions.

\subsubsection*{Ground States with $p\geq 2$}

We now turn to the ground states when the electrons carry an internal spin. We anticipated above that the states will take a simpler form when $N$ is divisible by $p$. And, indeed, this is the case.

\vskip 5pt
\noindent
\underline{$N$ divisible by $p$}
\para

When $N$ is divisible by $p$, there is a unique ground state. This is an  $SU(p)$ singlet. To describe the construction of this state, we first group $p$ creation operators $\varphi_i^\dagger$ together to form the $SU(p)$ baryon operator
\be {\cal B}(r)^\dagger_{a_1\ldots a_p} = \epsilon^{i_1\ldots i_p} (Z^r\varphi)^\dagger_{i_1\,a_1} \ldots (Z^r\varphi)^\dagger_{i_p\,a_p}\nn\ee
This is a singlet under the $SU(p)$ global symmetry, but transforms in the $p^{\rm th}$ antisymmetric representation of the  $U(N)$ gauge symmetry.
To construct an $SU(N)$ singlet with the correct $U(1)$  charge \eqn{traceconstraint}, we make a ``baryon of baryons". The ground state is then
\be |{\rm ground}\rangle_k = \left[ \epsilon^{a_1\ldots a_N} {\cal B}(0)^\dagger_{a_1\ldots a_p}{\cal B}(1)^\dagger_{a_{p+1}\ldots a_{2p}}\ldots 
{\cal B}(N/p-1)^\dagger_{a_{N-p+1}\ldots a_{N}} \right]^{k}|0\rangle\label{ground1}\ee
This state has energy $ E = \frac{\omega k}{2B} \frac{N(N-p)}{p}$. This time the requirements of the $U(N)$ gauge invariance have resulted in interesting correlations between both position and spin degrees of freedom of the electrons. We will devote the rest of this section and the next to describing the structure of these states.

%
%
%

\vskip 5pt
\noindent
\underline{$N = q$ mod $p$}
\para

When $N$ is not divisible by $p$, the ground state is no longer a singlet under the global $SU(p)$ symmetry. We write $N=m p + q$ with $m,q \in {\bf Z}^+$.  One can check that the ground states are
\be |{\rm ground}\rangle_k &=& \prod_{l=1}^{k} \left[ \epsilon^{a_1\ldots a_N}   {\cal B}(0)^\dagger_{a_1\ldots a_p}  {\cal B}(1)^\dagger_{a_{p+1}\ldots a_{2p}}\ldots  
{\cal B}(m -1)^\dagger_{a_{N-p-q+1}\ldots a_{N-q}} \right.\nn\\   &&\ \ \ \ \ \ \ \ \ \left.\ \ \ \ \ \   (Z^{m} \varphi_{i_{(l,1)}})^\dagger_{a_{N-q+1}}\ldots
(Z^{m} \varphi_{i_{(l,q)}})^\dagger_{a_N}
\right]|0\rangle\nn\ee
where $i_{(l,\alpha)}$, with $l=1,\ldots k$ and $\alpha= 1,\ldots q$ are free indices labelling the degenerate ground states. These ground states transform in the  $k^{\rm th}$-fold symmetrisation of the $q^{\rm th}$ antisymmetric representation of $SU(p)$. In terms of Young diagrams, this is the representation
\be   q\left\{\begin{array}{c} \\ \\ \\ \end{array}\right. \!\!\!\overbrace{\raisebox{-4.9ex}{\epsfxsize=1.3in\epsfbox{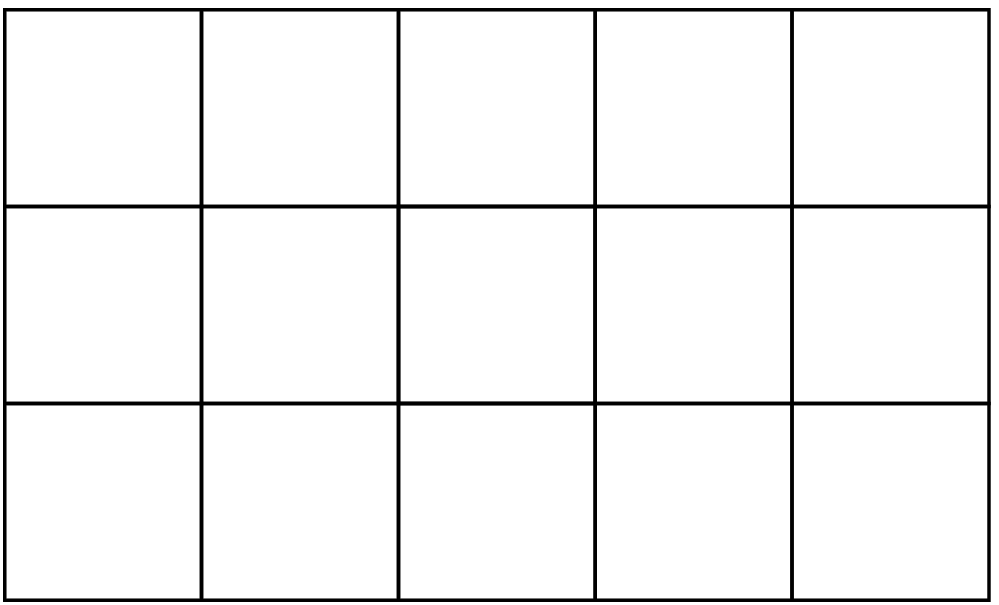}}}^k \nn\ee
We'll see in Section \ref{cftsec} why these representations are special and might be expected to arise in quantum Hall states. In the meantime, we will primarily focus on the states \eqn{ground1} that arise when $N$ is divisible by $p$.
%
%

\subsection{The Wavefunctions}

The description of the ground states given above is in terms of a coherent state representation for matrices. To make connections with the more traditional form of the wavefunctions, we need to find a map between the creation operators $Z^\dagger$ and the position space representation. For the $p=1$ states, this was explained by Karabali and Sakita \cite{ks,ks2}. We first review their results and then provide the generalisation to the $SU(p)$ matrix model.

\subsubsection*{$p=1$ and the Laughlin Wavefunctions}

At the formal level, there is a clear  similarity between the ground state for $p=1$ theories,
\be |{\rm ground}\rangle_k = \left[\epsilon^{a_1\ldots a_N} (Z^{0}\varphi)_{a_1}^\dagger (Z\varphi)^\dagger_{a_2}\ldots (Z^{N-1}\varphi)_{a_N}^\dagger\right]^k|0\rangle\label{pstate}\ee
%
%
%
%
and the Laughlin wavefunctions at filling fraction $\nu=1/m$
%
%
%
\be \psi^{\rm Laughlin}_{m}(z_a) &=& \prod_{a<b} (z_a-z_b)^m e^{-B\sum |z_a|^2/4} \nn\\ &=& \left[\epsilon^{a_1\ldots a_N}z_{a_1}^0z_{a_2}^1\ldots z_{a_N}^{N-1}\right]^m 
e^{-B\sum |z_a|^2/4} \nn\ee
However, this similarity can be misleading: the operators $Z^\dagger$ and $\varphi^\dagger$ are very different objects from the holomorphic position variables $z_a$. To make this connection precise, we need to be more careful about how to relate the two. In fact, there is no canonical map. There are, however, a number of natural ways to make the connection. Two of these, discussed in \cite{ks}, are:
\begin{itemize}
\item We work with a coherent state representation $\hat{Z}|Z,\varphi\rangle = Z|Z,\varphi\rangle$ and $\hat{\varphi}|Z,\varphi\rangle = \varphi|Z,\varphi\rangle$ where, for once, we've used hats to denote the difference between the quantum operator $\hat{Z}$ and the classical matrix $Z$. We then diagonalise $Z = VDV^{-1}$ with $D={\rm diag}(z_1,\ldots,z_N)$ and express the resulting wavefunctions as $\psi(z_a) = \langle z_a|\Psi\rangle$.
\item Alternatively, we could decompose the complex  operator matrix $\hat{Z}=\hat{X}+i\hat{Y}$ and subsequently work in a coherent state representation $\hat{X}|X\rangle = X|X\rangle$. This picture has the advantage that the matrices $\hat{X}$ and $\hat{Y}$ are conjugate, giving us the representation
\be  \hat{Z}^\dagger_{ab} = \frac{1}{\sqrt{2}} \left(X_{ab} - \frac \partial {\partial X_{ba}} \right) \nn \ee
Moreover, calculations in this approach are somewhat easier  because the diagonalisation $X=UXU^\dagger$ can be achieved by a unitary operator $U$. The resulting wavefunctions are written as $\psi(x_a) = \langle x_a|\Psi\rangle$. We then analytically continue $x_a\rightarrow z_a$ to provide holomorphic wavefunctions of the kind appropriate to describe the lowest Landau level. 
\end{itemize}
Both of these approaches were described in \cite{ks}. The resulting wavefunctions differ in detail, but share their most important properties. First, for $k=0$, the wavefunctions coincide with the Slater determinant for a fully-filled Landau level
\be \langle z_a\, |\,{\rm ground}\rangle_{k=0} = \prod_{a<b} (z_a-z_b) e^{-B\sum |z_a|^2/4}\label{groundzero}\ee
The exponential factor is the usual factor arising form the normalisation of coherent states.
The single factor of the Vandermonde determinant, which is not obvious in  \eqn{pstate} when $k=0$, is a Jacobian that arises in the transformation from matrix-valued objects to their eigenvalues.
 
 \para
 Meanwhile, for $k\geq 1$, neither representation of the wavefunction coincides with the Laughlin state. Nonetheless, both have the property that
 \be  \langle z_a \, |\,{\rm ground}\rangle_{k}\ \rightarrow\ \psi^{\rm Laughlin}_{k+1}(z_a) \ \ \ {\rm as}\ \ |z_a-z_b|\rightarrow \infty\nn\ee
In other words, the wavefunctions that arise from the matrix model coincide with the Laughlin wavefunctions only at large distances. This ensures that the matrix model ground state has filling fraction 
\be \nu = \frac{1}{k+1}\nn\ee
However, the matrix model state differs from the Laughlin wavefunction as particles approach to within a magnetic length. Indeed, one obtains the matrix model wavefunction from the Laughlin one by acting with exponentials of derivative operators $l_B (\partial/{\partial z})$ on the polynomial part. 
In particular, the matrix model wavefunctions do not exhibit the familiar zero-of-order $m$ that is characteristic of the Laughlin wavefunction. 

\para
The fact that  matrix model and Laughlin states differ in detail is not a matter of concern. There is nothing privileged about the Laughlin wavefunction: it is merely a representative of a universality class of states, characterised by their topological order. The matrix model state lies in the same universality class. This can be shown explicitly; for certain calculations the coherent state representation \eqn{pstate} offers greater analytic control. In particular, it was shown in \cite{us} that the fractional statistics of the quasi-hole can be computed in the coherent state representation following the classic calculation of \cite{arovas}, but without resorting to the plasma analogy.

\subsubsection*{Wavefunctions for $p\geq 2$}

For the case $p=1$ described above,  all physical states have the same dependence on  $\varphi^\dagger$ excitations; they differ only in their $Z^\dagger$ excitations. This is the reason that no $\varphi$ variables were needed when writing the wavefunctions.
In contrast, when $p\geq 2$, different physical states can have a different structure of $\varphi_i^\dagger$ excitations. These capture the way the state transforms under the $SU(p)$ symmetry.

\para
We repeat the procedure described above,  moving from coherent state representation to wavefunction. For $k=0$ the wavefunction knows nothing about the spin degrees of freedom. This means that the $k=0$ wavefunction is again given by \eqn{groundzero}, describing a fully-filled Landau level with $\nu = 1$. 

\para
However, for $p\geq 2$ and $k>1$, we have  a new ingredient.  Apart from the Vandermonde determinant \eqn{groundzero}, each time that a power of a particle coordinate $z_a$ appears in the wavefunction, it is accompanied by a spin degree of freedom, 
\be \sigma_a\ \in \{1,\ldots, p\}\nn\ee
where $a=1,\ldots, N$ labels the particle.

\para
For example, when $k=1$, each particle has a single spin degree of freedom $\sigma$. This reflects the fact that, as we saw earlier, each particle transforms in the fundamental representation of $SU(p)$. More generally, the internal state  of each particle is determined by $k$ factors of the spin label $\sigma$. As we will explain in some detail in Section \ref{bwsec}, this is to be interpreted as specifying the $k^{\rm th}$ symmetric representation under $SU(p)$.

\para
When $N$ is divisible by $p$, the ground state wavefunction \eqn{ground1} is a an  $SU(p)$ spin-singlet. The states have filling fraction 
\be \nu = \frac{p}{k+p}\label{fillingus}\ee
and have the property that
\be \langle z_a\,|\,{\rm ground}\rangle_k\ \rightarrow\ \psi_{BW}(z_a)  \ \ \ {\rm as}\ \ |z_a-z_b|\rightarrow \infty\nn\ee
where $\psi_{BW}(z_a)$ are a class of non-Abelian  wavefunctions constructed some time ago by Blok and Wen \cite{bw}. Like many non-Abelian quantum Hall states, the explicit description of the wavefunctions $\psi_{BW}(z_a)$ is straightforward, but somewhat fiddly.  We devote the next section to a more detailed description of these quantum Hall states and their properties.

\section{The Blok-Wen States}\label{bwsec}

In this section, we describe the Blok-Wen wavefunctions in some detail. The original construction of \cite{bw} was in terms of conformal blocks of a $SU(p)$ WZW model and we will revisit this approach in Section \ref{cftsec}. Here we provide an alternative, more down-to-earth construction of the states. We start with wavefunctions carrying spin under $SU(2)$, moving on to the more general $SU(p)$ case in Section \ref{supsec}. 

\subsection{Particles with $SU(2)$ Spin}\label{su2sec}

The simplest examples of wavefunctions describing particles with spin are due to Halperin  \cite{halperin}. We take $N$ particles, with $N$ even, and split them into two groups of $N/2$ particles, with positions $z_a$ and $w_a$ where each index now runs over  $a=1,\ldots, N/2$. The $(m,m,n)$ wavefunctions are
\be {\psi}(z,w) = \prod_{a<b}^{N/2}(z_a-z_b)^{m}\prod_{c<d}^{N/2}(w_c-w_d)^{m}\prod_{a,d} (z_a-w_d)^n\label{halperin}\ee
where, throughout this section, we will omit the overall exponential factor common to all wavefunctions. Counting the angular momentum of particles shows that these states have filling fraction 
%
\be \nu = \frac{2}{m+n}\nn\ee
The $(m,m,n)$ states \eqn{halperin} are really shorthand for wavefunctions with spin. As we review below, they should be dressed with explicit spin wavefunctions.  This will result in the Blok-Wen states. These are actually a slightly more general class of states than those that emerge from the matrix model. We will see that the matrix model gives states with $n=1$ and $m=k+1$.

\para
Usually one thinks of the Halperin states as describing spin-$\frac{1}{2}$ particles, with $z_a$ and $w_a$ labelling the positions of those which are spin-up and spin-down respectively. With this interpretation the  $(n+1,n+1,n)$ states are spin singlets. 
However, we will show that we can also view \eqn{halperin} as describing particles with  spin $s>\frac{1}{2}$. This is perhaps surprising as these particles have $2s+1$ spin states and it is not obvious how to decompose these into two groups. We will see that, with this interpretation, the $(m,m,n)$ states are spin singlets when $s=(m-n)/2$. Matching to the matrix model parameters, this means $s=k/2$.

\subsubsection*{Spin $\frac{1}{2}$}

The standard interpretation of \eqn{halperin} is as a wavefunction for spin-$\frac{1}{2}$ particles. To dress the wavefunction with these spin states, it's useful to change notation slightly and label the positions of all $N$ particles as $z_a$. Each particle carries a further internal spin degree of freedom $\sigma_a$ which takes values  $\up$ or $\down$. The $(m,m,n)$ state for $m>n$  is then written as
\be {\psi}(z,\sigma) = {\cal A}\left[\prod_{a<b}^{N}(z_a-z_b)^{n}  \!\!\! \prod_{a<b\ {\rm odd}} (z_a-z_b)^{m-n}  \!\!\!  \prod_{c<d\ {\rm even}} (z_c-z_d)^{m-n}\,\,|\uparrow\downarrow\uparrow\downarrow\ldots\uparrow\downarrow\rangle\right]\ \ \ \label{halpspin}\ee
%
%
%
where ${\cal A}$ stands for antisymmetrisation over all particles,  exchanging both positions and spins.  This wavefunction  describes fermions for $m$ odd and bosons for $m$ even.

\para
It is well known that only the states with $m-n=1$ are spin singlets \cite{haldane}. In this case, the wavefunction factorises as
\be \psi_{n+1,n+1,n}(z,\sigma) =  \prod_{a<b}^{N} (z_a-z_b)^n \ \Phi(z,\sigma)\nn\ee
This describes fermions for $n$ even and bosons for $n$ odd. 
Here the first factor takes the familiar Laughlin-Jastrow form, while the second factor is the Slater determinant of two fully filled Landau levels, one for the up spins and one for the down spins. The resulting wavefunction can be written as
\be \Phi(z,\sigma) &=& {\cal A}\Big[ \prod_{a<b\ {\rm odd}}(z_a-z_b) \prod_{c<d\ {\rm even}} (z_c-z_d)\ |\uparrow_1\rangle |\downarrow_2\rangle |\uparrow_3\rangle\ldots |\downarrow_{2N}\rangle\Big]\nn\ee
or, equivalently, as 
\be \Phi(z,\sigma) &=& \epsilon_{a_1\ldots a_{2N}} (z_{a_1} z_{a_2})^0 (z_{a_3} z_{a_4})^1 \ldots (z_{a_{2N-1}} z_{a_{2N}})^{N-1} \nn\\ 
&& \ \ \ \ \ \ \times\ \Big[  |\uparrow_{a_1}\rangle |\downarrow_{a_2}\rangle |\uparrow_{a_3} \rangle| \downarrow_{a_4}\rangle  \ldots  |\uparrow_{a_{2N-1}}\rangle | \downarrow_{a_{2N}}\rangle \Big]\label{phi}\ee
In particular, this latter form makes it clear that the spins are paired in singlet states of the form $|\uparrow_{a_1}\rangle |\downarrow_{a_2}\rangle - |\downarrow_{a_1}\rangle |\uparrow_{a_2}\rangle$.

\subsubsection*{Spin 1}

So far, we have just reproduced the usual story. Suppose now that $m=n+2$. We claim that the following is a spin-singlet wavefunction for spin 1 particles,
\be
\psi_{n+2,n+2,n}(z,\sigma) =  \prod_{a<b} (z_a-z_b)^n \ {\cal P}\left[\Phi^2(z,\sigma)\right]\label{n2}\ee
This is a wavefunction for fermions when $n$ is odd and bosons when $n$ is even. 

\para
Our first  task is to explain what this means. The factor $\Phi^2$ includes two spin states for each particle. 
The tensor product of two spin 1/2 states gives ${\bf 2}\otimes {\bf 2} = {\bf 1} \oplus {\bf 3}$. The operator ${\cal P}$ projects onto the symmetric ${\bf 3}$. (In the present case, this operation is not required as it is implemented automatically by the form of $\Phi^2$. However, we include it in our expression for clarity.) This means that we can interpret \eqn{n2} as a quantum Hall state for spin 1 particles, with the map 
\be \up\up = \ket{1}\ \ \ ,\ \ \ \down\down = \ket{-1}\ \ \ ,\ \ \ \up\down=\down\up = \ket{0}\label{12map}\ee
 We further claim that \eqn{n2} is a spin singlet. We will first motivate this by looking at the kinds of terms that arise. We will subsequently provide a proof in the course of rewriting the wavefunction in a more familiar form.

\para
Consider two particles, labelled 1 and 2, each of which carries spin $\frac{1}{2}$. The spin singlet state is
\be \ket{12}_{\frac{1}{2}} =  \ket{\uparrow_1}\ket{\downarrow_2} - \ket{\downarrow_1}\ket{\uparrow_2}\nn\ee
where the subscript $1/2$ is there to remind us that this is the singlet built from two spin 1/2 particles. The simplest terms that occur in \eqn{n2} are of the form $\ket{12}_{\frac{1}{2}}\ket{12}_{\frac{1}{2}}$. Using the map \eqn{12map}, we have
\be \ket{12}_{\frac{1}{2}}\ket{12}_{\frac{1}{2}} &=& \ket{1_1}\ket{-1_2} + \ket{-1_1}\ket{1_2} - 2\ket{0_1}\ket{0_2} \nn\ee
which is indeed the singlet formed from two spin 1 states. To highlight this, we write the above equation as
%
%
%
\be \ket{12}_{\frac{1}{2}}\ket{12}_{\frac{1}{2}}= \ket{12}_1\nn\ee
The next kind of term that arises in \eqn{n2} involves four different particles. It is $\ket{12}_{\frac{1}{2}}\ket{23}_{\frac{1}{2}}\ket{34}_{\frac{1}{2}}\ket{41}_{\frac{1}{2}}$. We can similarly expand this in terms of spin 1 states and again find that only combinations of singlet states appear:
\be \ket{12}_{\frac{1}{2}}\ket{23}_{\frac{1}{2}}\ket{34}_{\frac{1}{2}}\ket{41}_{\frac{1}{2}}  = \ket{12}_1\ket{34}_1 - \ket{13}_1\ket{24}_1 + \ket{14}_1\ket{23}_1\nn\ee
The most general term in \eqn{n2}  has $2n$ particles. This too can be written as the linear combinations of $n$ spin 1 singlet states.  Rather than demonstrate this term by term, we will instead  show that the wavefunction \eqn{n2} has an alternative form written purely in terms of spin 1 singlets.

\subsubsection*{The Spin 1 Wavefunction as a Pfaffian}

We will now show that the  wavefunction  \eqn{n2} for spin 1 particles  can be written as
\be \Phi^2(z,\sigma) = {\rm Pf}\left(\frac{|ab\rangle_1}{z_a-z_b}\right)\prod_{a<b}(z_a-z_b)\label{pfaffian}\ee
%
%
%
%
with ${\rm Pf}(M_{ab})$ the Pfaffian of the matrix $M$. This is a spin singlet version of the Moore-Read state \cite{mr}.
It is sensible because the spin 1 singlet $|ab\rangle_1$ is symmetric in the two spins, in contrast to $|ab\rangle_{\frac{1}{2}}$ which is antisymmetric. 

\para
It was noticed long ago \cite{bw,ho}  that the $(3,3,1)$ state  is closely related to the Pfaffian state.  In \cite{bw} the particles were spin-1 but projected onto the $m=0$ spin component; in \cite{ho} the particles were taken to be
 spin 1/2 and the resulting state was not a spin singlet. Our result \eqn{pfaffian} is clearly closely related to these earlier results, both of which are proven using the Cauchy identity. However, the proof of \eqn{pfaffian} requires more sophisticated machinery which appears not to have been available at the time of \cite{bw,ho}.

\para
\underline{The Proof:}

\para
The projective Hilbert space associated to the two spins is a Bloch sphere ${\bf CP}^1$. We parameterise this by the inhomogeneous coordinate $\zeta$.  Formally, we then  set $|\!\downarrow_a\rangle = 1$ and $|\!\uparrow_a\rangle = \zeta_a$ and write $\Phi$ as the polynomial
\be \Phi(z,\zeta) = 
\frac{1}{2^{N/2}}\epsilon_{a_1\ldots a_{N}}\Big[(z_{a_1}z_{a_2})^0\ldots (z_{a_{N-1}}z_{a_{N}})^{N/2-1}\Big] \Big[(\zeta_{a_1}-\zeta_{a_2})\ldots (\zeta_{a_{N-1}} - \zeta_{a_{N}})\Big]\nn\ee
This has the advantage that the right-hand-side can be viewed as the determinant of a $N\times N$ matrix $\Delta[z;\zeta]$ with components given by
\be \Delta[z;\zeta]_{a,b} = \left\{\begin{array}{lc} \ \ z_a^{b-1} \ \ \ \ \ \ \ \ \ \ & 1\leq j \leq \frac{N}{2}\\ \ \zeta_a z_a^{b-1} & \frac{N}{2}+1\leq j \leq 2N\end{array}\right.\label{delta}\ee
To show the result \eqn{pfaffian}, we then need to prove the polynomial identity
\be {\rm det}^2\Delta[z;\zeta]\ \stackrel{?}{=}\  {\rm Pf}\left(\frac{(\zeta_a-\zeta_b)^2}{z_a-z_b}\right)\prod_{a<b}(z_a-z_b)\nn\ee
%
%
%
In fact, this identity is a special case of a more general result proven in \cite{okado}. Theorem 2.4 of this paper shows (among other things) that two matrices $\Delta[z;\zeta]$ and $\Delta[z;\eta]$, each  defined by \eqn{delta}, obey the relation
\be  {\rm det}\Delta[z;\zeta]\,{\rm det}\Delta[z;\eta] = {\rm Pf}\left(\frac{(\zeta_b-\zeta_a)(\eta_b-\eta_a)}{z_b-z_a}\right) {\rm det}(z_a^{b-1})  \nn\ee
Setting $\zeta_a = \eta_a$ yields the desired result. \hfill$\Box$

%
%
%
%

\subsubsection*{Higher Spin}\label{highspinsec}

The generalisation to higher spins is now obvious. We construct the wavefunction
\be \psi_{n+2s,n+2s,n}(z,\sigma) =  \prod_{a<b} (z_a-z_b)^n \ {\cal P}\left[ \Phi^{2s}(z,\sigma)\right]\label{n3}\ee
where ${\cal P}$ is there to remind us that the spin states for each particle are projected onto the fully symmetrised product. 
This means that this is a wavefunction for particles with spin $s$. Once again, the final state is a spin singlet. This follows from some trivial group theory. The infinitesimal action of $SU(2)$ on the tensor product of $2s$ spin states is
\be T^\alpha = \sum_{a=1}^N t^\alpha_a \otimes {\bf 1}\otimes \ldots \otimes {\bf 1} + \mbox{symmetric}\nn\ee
where $t^\alpha_a$ is the operator in the fundamental representation acting on the $a^{\rm th}$ particle, and $\alpha=1,2,3$ labels the three $su(2)$ generators. Because ${\cal P}$ projects onto an irrep, we have
\be T^\alpha {\cal P}\left[\Phi^2s\right] = {\cal P}\left[\sum_a t^\alpha_a \Phi\otimes \Phi\otimes \ldots \otimes \Phi\right] + \mbox{symmetric}\nn\ee
But each of these terms vanishes because $\Phi$ is itself a spin-singlet, which means that $\sum_a t^\alpha_a\Phi=0$. This ensures that \eqn{n3} is indeed a spin singlet.

\para
Although \eqn{n3} provides an explicit description of the state, it would be pleasing to find a simple expression purely in terms of the singlets $|ab\rangle_s$, analogous to the Pfaffian \eqn{pfaffian} for $s=1$. We have not been able to do this; it may simply not be possible due to the entanglement structure between higher numbers of spins.

\para
While the Halperin states \eqn{halperin} describe Abelian quantum Hall states, our spin singlet states \eqn{pfaffian} and \eqn{n3} with spin $s\geq 1$ are all non-Abelian quantum Hall states. Indeed, it has long been known  that dressing a quantum Hall state with spin degrees of freedom can change the universality class of the state. We will see in Section \ref{cftsec} that these states are associated to $SU(2)_{2s}$ WZW models.

\subsection{Particles with $SU(p)$ Spin}\label{supsec}

We now generalise these ideas to particles that carry a ``spin" under the group $SU(p)$. This means that  each particle carries an internal Hilbert space which transforms under a particular representation of $SU(p)$. 

\para
The starting point is the $p$-component generalisation of the Halperin states \eqn{halperin}. We take $N$ particles and split them into $p$ groups, with positions $w_{i\,a}$, where $i=1,\ldots,p$ and $a=1,\ldots, N/p$.  Then 
\be {\psi}_{m,n}(z) = \left[\prod_{i=1}^p\prod_{a<b}^{N/p}(w_{i\,a}-w_{i\,b})^{m}\right] \left[\prod_{i<j}^p\prod_{c,d}^{N/p} (w_{i\,c}-w_{j\,d})^n\right]\label{multihalp}\ee
Multi-component states of this form were first discussed in \cite{layer}. More recently they have been studied  in \cite{goer1} for $p=4$ to describe both spin and valley indices of electrons in graphene, and more generally in \cite{goer2,goer3}. The states \eqn{multihalp} have filling fraction
\be \nu = \frac{p}{pn +(m-n)}\label{goodfilling}\ee
It is  natural to think of these wavefunctions as describing objects with $p$ internal states. This corresponds to the situation where each particle sits in the fundamental representation, ${\bf p}$ of $SU(p)$. However, as we will see, there is also a  generalisation of our previous construction in which each particle has more internal states, corresponding to the symmetric representations  of $SU(p)$.

\subsubsection*{Fundamental Representation}

We  start by describing the simplest situation where the particles sit in the fundamental representation, meaning that each carries an internal  index, $\sigma_a \in \{1,2,\ldots,p\}$. In this case, the wavefunctions \eqn{multihalp} are spin-singlets when $m=n+1$. 

\para
To see this, note that the smallest number of particles that can form a singlet state is $p$. To achieve this, the spin degrees of freedom are completely antisymmetrised in what  high energy physicists would call a ``baryon",
\be B_{a_1\ldots a_p} = \epsilon^{\sigma_{a_1}\ldots \sigma_{a_p}} \ket{\sigma_{a_1}}\ldots \ket{\sigma_{a_p}}\nn\ee
We can then form a spin singlet state of type $(n+1,n)$ by writing
\be \psi_{n+1,n}(z,\sigma) = \prod_{a<b}^{N} (z_a-z_b)^n \Phi_{(p)}(z,\sigma)\nn\ee
where, in analogy with \eqn{phi}, $\Phi_{(p)}$ describes $p$ fully filled Landau levels, one for each type of spin,
\be \Phi_{(p)}(z,\sigma) &=& \epsilon_{a_1\ldots a_{N}} (z_{a_1} \ldots z_{a_p})^0 (z_{a_{p+1}} \ldots z_{a_{2p}})^1 \ldots (z_{a_{N-p+1}}\ldots z_{a_{N}})^{N-1} \nn\\ 
&& \ \ \ \ \ \ \times\ B_{a_1\ldots a_p}B_{a_{p+1}\ldots a_{2p}}\ldots B_{a_{N-p+1}\ldots a_{N}}\label{phip}\ee
In the language of \cite{rr,ardschout}, this state exhibits clustering at order $p$. This means that the factor $\Phi_{(p)}$ remains non-zero if the positions of up to $p$ particles coincide. However, it vanishes if $p+1$ or more particles coincide. We will see in Section \ref{cftsec} that this wavefunction actually describes an Abelian quantum Hall state. To generate non-Abelian quantum Hall states, we need to look at higher representations of $SU(p)$.

\subsubsection*{Symmetric Representations}

For $m>n+1$, we can still interpret \eqn{multihalp} as a spin-singlet state, but now each particle must carry a spin in a higher representation. We define $k=m-n$ and write the wavefunction as
\be \psi_{n+k,n}(z,\sigma) = \prod_{a<b}^{N} (z_a-z_b)^n\, {\cal P}\left[ \Phi^k_{(p)}(z,\sigma)\right]\label{final}\ee
where ${\cal P}$ projects onto the symmetrised product of spin states, meaning that each particle transforms in the  $k^{\rm th}$ symmetric representation of $SU(p)$. These states are all spin singlets, by the same argument that we gave in Section \ref{highspinsec}.

\para
These states, still with $n = 1$, exactly reproduce the long-distance behaviour of the matrix model ground states described in the previous section, with the same values of $k$ and $p$.


\subsubsection*{Relationship to  Read-Rezayi States}

When $k=2$, our states describe particles transforming in the symmetric representation of $SU(p)$ with dimension $\frac{1}{2}p(p+1)$. They  are $p$-clustered states with filling fraction 
\be \nu = \frac{p}{pn+2}\nn\ee
Both of these properties are shared by the Read-Rezayi states \cite{rr}. We will now show that our states are spin-singlet generalisations of the Read-Rezayi states.

\para
A particularly simple form of the Read-Rezayi state was presented in \cite{capelli},
\be \psi^{RR}_n(z) = {\cal S} \left[\prod_{i=1}^p\prod_{a<b}^{N/p}(w_{i\,a}-w_{i\,b})^2\right]\prod_{c<d}^{N} (z_c-z_d)^{n}\label{crr}\ee
where ${\cal S}$ means that we symmetrise over all possible divisions of the particles into the $p$ groups, while the $z_a$ factor simply means that we include all particle positions rather than restricting to those in a specific group.

\para
We will show that, after projection onto a particular spin state, the Blok-Wen wavefunction \eqn{final} coincides with the Read-Rezayi wavefunction \eqn{crr}, i.e. 
\be {\cal P}\left[ \Phi^k_{(p)}(z,\sigma)\right]\ \mapsto\ {\cal S} \left[\prod_{i=1}^p\prod_{a<b}^{N/p}(w_{i\,a}-w_{i\,b})^2\right]\label{mapsto}\ee

\para
\underline{The Proof:}

\para
Our first task is to explain what $\mapsto$ means in the above expression.  It's useful to first revisit the case of the Pfaffian \eqn{pfaffian}. There, the spin singlet wavefunction included the factor
\be \ket{ab}_1 = \ket{1_a}\ket{-1_b} + \ket{-1_a}\ket{1_b} - 2\ket{0_a}\ket{0_b} \nn\ee
and we can project onto a spin-polarised state simply by replacing $|ab\rangle_1 \mapsto |0_a\rangle|0_b\rangle$. This point was also made in \cite{bw}. 

\para
Unfortunately, there is no analogous procedure for $SU(p)$ spins. There is, however, a generalisation of the projection onto the highest spin states. The states $\ket{1_a} = \ket{\uparrow_a}\ket{\uparrow_a}$ and $\ket{-1_a} = \ket{\downarrow_a}\ket{\downarrow_a}$ have the property that both fundamental spins lie in the same direction. This is something which also makes sense for $SU(p)$ spins. We therefore define the projection $\mapsto$ in \eqn{crr} as an operator which correlates the fundamental $SU(p)$ spins associated to each individual particle
\be  \prod_{a=1}^N \ket{\sigma_a}\ket{\sigma'_a}  \ \mapsto \prod_{a=1}^N \delta_{\sigma_a \sigma'_a} \nn\ee
In particular, when we project the state ${\cal P}\left[ \Phi^2_{(p)}(z,\sigma)\right]$, we  correlate the two antisymmetrisations of spins  inside the two $\Phi_{(p)}$ factors. The projection picks out the states in which these two spins associated to a given particle are the same. Spins associated to different particles can be different.


%


\def\rrproj{\mathcal{P}_S}

%
%
%

\para
Having defined the projection, we turn to the structure of the state \eqn{final}. 
 It is helpful to think about collecting terms with some particular allocation of spin to each particle. For definiteness, let us consider the term where the particle at position $z_{(a-1)p+i} \equiv w_{i\,a}$ is given spin $i$. (Here $a = 1, \ldots,N/p$.) Now consider the polynomial in $z$ which multiplies this spin state. It is the antisymmetrisation over all ways of permuting particles at positions $w_{i\,a}$ and $w_{i\,a'}$ of 
\be (w_{1\,1} \ldots w_{p\,1})^0 (w_{1\,2} \ldots w_{p\,2})^1 \cdots (w_{1\,N/p} \ldots w_{p\,N/p})^{N/p-1} \nn\ee
But this is simply the product of $p$ separate Laughlin-like factors $w_{a\,1}^0 w_{a\,2}^1 \cdots w_{a\,N/p}^{N/p-1}$, and hence the coefficient of the spin state is proportional to
\be \prod_{a=1}^p\prod_{a<b}^{N/p}(w_{ai}-w_{aj}) \label{genhalperincoeff} \ee
which is essentially half of the Read-Rezayi state. To complete the argument, notice that whatever sign a particular spin allocation comes with, it comes with the same sign in both copies of $\Phi_{(p)}$. Hence overall, we obtain the square of this expression, \emph{symmetrised} over all spin allocations.  The projection does indeed result in the Read-Rezayi state \eqn{mapsto}.\hfill$\Box$

\para
The  Read-Rezayi states are associated to the parafermion CFT $SU(2)_k/U(1)_k$. Meanwhile, our states are associated to $SU(k)_2$. The two are related by level-rank duality. This means that our states include the non-Abelian anyons of the Read-Rezayi state; for example, $SU(3)_2$ includes the Fibonacci anyons. We will see how these emerge in Section \ref{cftsec} when we review the connection to conformal field theory. However, the Blok-Wen spin-singlet states arise from a CFT with no quotient, and hence contain additional anyonic degrees of freedom that are not part of the Read-Rezayi sequence of states.

\section{The View from Chern-Simons Theory}\label{vortexsec}

Until now, we've focussed only on the properties of the matrix model \eqn{matrix}. In this section, we explain where it comes from. The main idea, first proposed in \cite{unknown} and recently explored in some detail in \cite{us}, is that the matrix model describes the dynamics of vortices in a $d=2+1$ dimensional Chern-Simons theory. In \cite{us}, this connection was explained for Abelian Chern-Simons theories; in this section we generalise this picture to the non-Abelian case.

\para
Our starting point is a Chern-Simons theory with gauge group 
\be U(p)_{k',k} = \frac{U(1)_{k'} \times SU(p)_k}{{\bf Z}_p}\label{upgauge}\ee
The ${\bf Z}_p$ quotient places a strong restriction on the allowed values of $k'$ which must obey \cite{sw}
\be k'-kp\in p^2 {\bf Z}\label{sw}\ee
We denote the $U(1)$ gauge field as $\tilde{a}$ and the $SU(p)$ gauge field as $a$. Both are to be thought of as emergent gauge fields in the condensed matter system. Their dynamics is governed by the Chern-Simons action
\be S_{CS} = -\int d^3x\  \frac{k'}{4\pi} \epsilon^{\mu\nu\rho} \tilde{a}_\mu\partial_\nu\tilde{a}_\rho + \frac{k}{4\pi}{\rm Tr}\,\epsilon^{\mu\nu\rho}(a_\mu \partial_\nu a_\rho - \frac{2i}{3} a_\mu a_\nu a_\rho)\nn\ee
To this we couple non-relativistic matter. We consider $N_f$ bosons $\phi_i$, with $i=1,\ldots, N_f$, each transforming in the ${\bf p}$ of $SU(p)$,  with charge 1 under the $U(1)$. Their  action is 
%
%
%
%
\be S_{\rm matter} = \int d^3x\   i\phi_i^\dagger {\cal D}_0 \phi_i  - \,\frac{1}{2m} {\cal D}_n \phi_i^\dagger{\cal D}_n \phi_i 
    -\, \frac{\pi}{mk'}(\phi_i^\dagger\phi_i)^2 - \frac{\pi}{mk} (\phi_i^\dagger t^\alpha\phi_i)^2 \nn \ee
Here the subscripts $\mu,\nu,\rho=0,1,2$ are spacetime indices while $n=1,2$ is a spatial index only.  The $SU(p)$ generators $t^\alpha$ are in the fundamental representation. 

\para 
The coefficients of the $\phi^4$ terms --- which each describe the strength of a delta-function interaction between particles --- are seemingly fine-tuned to be proportional to the Chern-Simons levels $1/k'$ and $1/k$. However, these coefficient are known to run logarithmically under RG flow and we have simply set them to their fixed points \cite{bergman,bbak}. As an alternative justification, one could add fermions to this theory  and, with this choice of the coefficient, complete  it in a manner consistent with supersymmetry \cite{llm}. 

\para
The full action is then
\be S_{\rm 3d} =  S_{CS} + S_{{\rm matter}} - \int d^3x \ \mu \tilde{a}_0   \nn \ee
where we've introduced a background charge  $\mu$. This causes the scalars to condense in the vacuum, breaking the gauge symmetry. This symmetry breaking is complete whenever $N_f\geq p$. 
 In what follows we will take $N_f=p$. There is a unique ground state of the theory given by
\be \phi_i^a = \sqrt{\frac{\mu}{p}} \,\delta_i^a\nn\ee
with $a=1,\ldots,p$ the gauge index and $i=1,\ldots,p$ the flavour index.  In this vacuum, the gauge and flavour symmetries are broken to 
\be U(1)_{\rm gauge} \times SU(p)_{\rm gauge}\times SU(p)_{\rm flavour}\  \longrightarrow \ SU(p)_{\rm diag}\label{breaking}\ee
The low-energy physics of this broken phase is {\it not} that of a quantum Hall fluid. However, this can change in the presence of vortices.

\subsubsection*{Vortices}
 
 The symmetry breaking pattern \eqn{breaking} allows for the existence of vortex excitations in which the phase of $\phi$ winds. These have a rather nice property in this theory. The fine-tuning of the potential term described above means that vortices lie at the ``Bogomolnyi point"; they satisfy first-order differential equations, rather than second order ones. The vortex equations are
\be \tilde{f}_{12} = \frac{2\pi}{k'}\left(|\phi_i|^2 - \mu\right)\ \ \ ,\ \ \ f^\alpha_{12} = \frac{2\pi}{k}\phi_i^\dagger t^\alpha \phi_i \ \ \ ,\ \ \ {\cal D}_z\phi_i=0\label{vortexbog}\ee
where the first and second  of these equations are Gauss' law for the Abelian and non-Abelian gauge field respectively. In the Abelian case, BPS vortices also appeared in the context of quantum Hall physics in \cite{typology}. 

\para
These equations coincide with the vortex equations that arise in certain non-Abelian relativistic theories \cite{me,yung}. Their properties have been studied in some detail over the years (see, for example, \cite{nitta,shifyung,mereview}), especially in the case $k'=kp$, which is consistent with \eqn{sw}, where $U(1)$ and $SU(p)$ gauge fields naturally combine into a $U(p)$ gauge fields with the same level, meaning that  Gauss' law reads
\be f_{12} = \frac{2\pi}{k}\left(\phi_i\phi_i^\dagger - \frac{\mu}{p}\right)\nn\ee
The most striking fact about these equations is that they do not have a unique solution. Instead, for $N$ vortices the most general solution has $2pN$ parameters \cite{me}. (This is shown by index theorem techniques, generalising previous results for Abelian vortices \cite{erick,taubes}.) These parameters can be thought of as labelling the positions and internal orientations of the $N$ vortices. In particular, there are no forces between vortices. They can sit anywhere on the plane. 

\para
We can pick out a unique solution by adding an external harmonic trap. We choose a trap which, when evaluated on vortices, is proportional to their angular momentum
\be V_{\rm trap} = -\omega\int d^2x\ \frac{\mu|z|^2}{2p} \tilde{f}_{12}\nn\ee
This, of course, changes the equations of motion. The previous vortex solutions now precess around the origin. There is a unique, stationary, lowest-energy state which occurs  when all vortices coalesce at the origin to form a rotationally invariant configuration.

\para
For a large number of vortices $N$ the solution looks like a disc of radius
\be R \approx \sqrt{\frac{k' N}{\pi\mu p}}\nn\ee
Inside this disc, the scalar fields vanish, $\phi\approx 0$ and Gauss' law is satisfied by the presence of a constant magnetic flux $\tilde{f}_{12} \approx -2\pi\mu/k'$. 
%
%
The end result is that we have manufactured a disc shaped region, inside of which lives an unbroken Chern-Simons theory with $U(p)$ gauge group \eqn{upgauge}. We view this as a region of quantum Hall fluid.

\subsubsection*{Quantising the Vortices}

We can start with a few simple observations. Our $d=2+1$ dimensional theory has a background charge density $\mu$. From the perspective of the vortices, this looks like an effective external magnetic field \cite{us},
\be B = \frac{2\pi\mu}{p}\ee
With this information, we can estimate the filling fraction of the quantum Hall fluid of vortices. In an area $A = \pi R^2$, the number of states in the lowest Landau level is $BA/2\pi =  k' N/p^2$. Since we have filled this disc with $N$ vortices, we expect a filling fraction
\be \nu = \frac{p^2}{k'}\label{vortexfill}\ee
To understand the detailed properties of this Hall fluid, we must look in more detail at the microscopic dynamics of the vortices. 
As we mentioned above, in the absence of a trap, the most general solution to the vortex equations \eqn{vortexbog} has $2pN$ collective coordinates. We think of these as parameterising a manifold ${\cal M}_{p,N}$ which is called the {\it moduli space}. Each point on the moduli space corresponds to a different solution. Because the action is first-order in time derivatives, rather than second-order, the moduli space ${\cal M}_{p,N}$ should be thought of as the {\it phase space} of vortices rather than the configuration space. 

\para
An expression for the first-order dynamics of these vortices was derived in \cite{manton}. (See also \cite{romao,nunomartin,us}; these papers all deal with vortices in the Abelian theory, but the generalisation to non-Abelian vortices is straightforward.) Unfortunately, this result is somewhat abstract and, for a large number of closely packed vortices, not particularly useful. 

\para
Instead, we turn to a more versatile construction of the non-Abelian vortex moduli space ${\cal M}_{p,N}$ first derived in \cite{me} using D-brane techniques. We introduce a complex $N\times N$ matrix $Z$ and $p$ complex $N$-vectors $\varphi_i$, $i=1,\ldots,p$. Then the vortex moduli space is isomorphic to the space of solutions to 
\be \frac{B}{2}[Z,Z^\dagger] +  \sum_{i=1}^p\vphi_i\vphi_i^\dagger = k'{\bf 1}_N\nn\ee
with solutions identified if they are related by $Z\rightarrow U Z U^\dagger$ and $\varphi_i\rightarrow U\varphi_i$ where $U\in U(N)$. This, of course, is precisely the phase space of the matrix model \eqn{matrix}, with the constraint above arising as Gauss' law \eqn{constraint}. Moreover, the dynamics of the matrix model coincides with the dynamics expected on the vortex moduli space. 

\para
The phase space of the matrix model and the vortex moduli space ${\cal M}_{p,N}$ are believed to coincide  as complex manifolds, with the same K\"ahler class. However, the symplectic form  on the phase space inherited from the quotient construction does not coincide with that associated to vortex dynamics. This means that the matrix model should be used with some caution in extracting detailed properties of the vortices. However, our interest lies in the universality class of the quantum Hall ground states and here the matrix model is expected to give the right answer. Indeed, we've seen that the ground state of the matrix model lies in the same universality class as the Blok-Wen states. In the next section, we will confirm that these Blok-Wen states are indeed the ground state wavefunctions associated to the Chern-Simons theory \eqn{upgauge}.

\para
There is one final subtlety. Classically, the matrix model describes the dynamics of vortices when the $U(1)$ and $SU(p)$ levels are equal: $k'=kp$. However, at the quantum level, there is a shift of the level. In the 3d Chern-Simons theory, the $SU(p)$ level is renormalised at one-loop to $k\rightarrow k+p$. The matrix model captures the quantum dynamics when these shifted levels coincide. This requires
\be k'=(k+p)p\nn\ee
This too satisfies the requirement \eqn{sw}. This is the value that we've used in \eqn{matrix} and throughout this paper. 
In particular, we see that the filling fraction \eqn{vortexfill} becomes $\nu = p/(k+p)$ in agreement with the matrix model result \eqn{fillingus}.

\section{The View from Conformal Field Theory}\label{cftsec}

In the previous section, we used vortices to construct  a disc-like region of space  in which the low-energy dynamics is described by an unbroken $U(p)$ Chern-Simons theory. The microscopic dynamics of these vortices are described by the matrix model \eqn{matrix} whose ground states lie in the same universality class as the Blok-Wen wavefunctions. In this section, we close the circle and describe these states from the perspective of the boundary.

\para
Our vortex construction has presented us with a Chern-Simons theory on a manifold with boundary, where the boundary is now the edge of the large vortex.  On general grounds, we expect this boundary to support a chiral $U_{(k+p)p,k}$ WZW model \cite{wittenknot,israeli}. This should manifest itself in two ways. First, the excitations of the matrix model should coincide with the excitations of a (suitably discretised) WZW model. We will return to this in future work \cite{later}. Secondly, the ground state wavefunction --- which, as we have seen, is of the Blok-Wen type --- should arise as the correlation function in the conformal field theory \cite{mr}. This, of course, was how Blok and Wen originally derived their wavefunctions \cite{bw}. Here we review this construction, including the effect of the Abelian factor in the gauge group.

\para
Let's first review some simple properties of the WZW models. The irreducible representations of the $SU(p)$ Kac-Moody algebra at level $k$ are labelled by the corresponding representation of the $SU(p)$ Lie algebra. The latter are well known to be described by Young tableaux with up to $p-1$ rows. The representations of $SU(p)_k$ are those Young tableaux which have no more than $k$ boxes in the first row.

\para
Each irreducible representation of the Kac-Moody algebra gives rise to a primary operator in the corresponding WZW model. We call these operators ${\cal O}_{R}$ where $R$ denotes the representation. The usual candidates for quantum Hall wavefunctions are the correlation functions of strings of chiral operators
\be \langle {\cal O}_R(z_1)\ldots {\cal O}_R(z_{N})\rangle\label{string}\ee
where ${\cal O}_R$ is the ``electron operator" in the CFT or, more generally, the operator associated to the particle which forms the quantum Hall state. 
(More precisely, these should be thought of as conformal blocks of the non-chiral WZW theory.)

\para
There is, however, a problem in identifying \eqn{string} as a wavefunction: for most choices of ${\cal O}_R$, there is no unique answer due to monodromies in the correlation function as $z_a$ are varied. Instead, the  number of conformal blocks is the number of singlets that arises when the many copies of $R$ are fused together. 
Typically this number will increase exponentially with $N$. Of course, this growth of conformal blocks is precisely what's needed to describe non-Abelian quasi-holes in a quantum Hall state, but this should only occur for correlation functions in which quasi-hole operators are inserted. For a sensible quantum Hall interpretation, we want to have a unique ground state, and this means that \eqn{string} should yield a unique answer when only electron operators are inserted.

\para
There is, fortunately, a choice of $R$ for which \eqn{string} has a unique answer. We take $N$ to be a multiple of $p$ and choose  the representation $R$ which is maximally symmetric. In terms of Young diagrams, it is a single row of $k$ boxes
\be R = \ \  \overbrace{\raisebox{-1.2ex}{\epsfxsize=1.3in\epsfbox{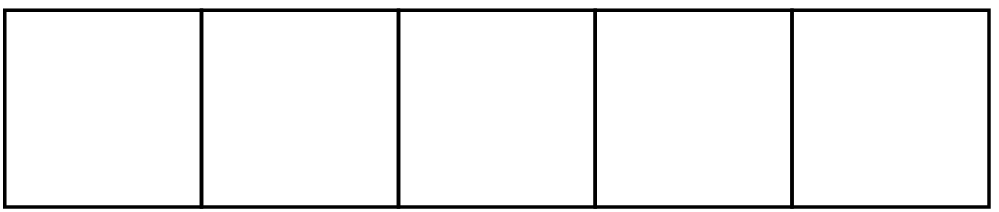}}}^k\label{r}\ee
For $SU(2)_k$, this corresponds to the spin $s=k/2$ representation; for $SU(p)_k$ it is the $k^{\rm th}$ symmetric representation. This, of course, is precisely the representation carried by the particles described by our wavefunctions \eqn{final}.

\para
To see that there is indeed a unique singlet when we take \eqn{r}, we need to look at the fusion rules. (See, for example \cite{yellow}, for a detailed description of how these are computed.\footnote{A Mathematica code for computing  $SU(p)_k$ fusion rules, written by one of us (CT), can be downloaded at \href{http://blog.suchideas.com/2mBUW}{http://blog.suchideas.com/2mBUW}.})
For $SU(2)_k$, it is straightforward to show that the fusion of two spin $s=k/2$ representations leaves only the singlet $s=0$. Written in terms of the dimension ${\bf d} = 2s+1$ of the representation, this reads
\be  ({\bf k +1})\star ({\bf k+1}) = {\bf 1}\nn\ee
For $SU(3)_k$, one finds that the $k^{\rm th}$ symmetric representation, which we denote  ${\rm Sym}_k$, has fusion rules
\be {\rm Sym}_k\star {\rm Sym}_k = \overline{\rm Sym}_k\nn\ee
while 
\be {\rm Sym}_k \star \overline{\rm Sym}_k = {\bf 1}\nn\ee
which tells us that three copies of ${\rm Sym}_k$ can fuse only to the singlet. 
More generally, for $SU(p)_k$ we define the representations 
\be Y_i = \ i\left\{\begin{array}{c} \\ \\ \\ \end{array}\right. \!\!\!\overbrace{\raisebox{-4.9ex}{\epsfxsize=1.3in\epsfbox{young2.eps}}}^k \nn\ee
which consist of $k$ columns (the maximum number allowed by the level) and $i$ rows. These are precisely the representations that we saw in Section \ref{groundsec} when discussing  the ground states of the matrix model.  In this notation, our maximally symmetric representation is $Y_1 = {\rm Sym}_k$, while $Y_{N-1} = \overline{\rm Sym}_k$. The $Y_i$ have the nice property that they fuse only among themselves \cite{sy}
\be Y_i\star Y_j = Y_{i+j\ {\rm mod}\ p}\nn\ee
This is enough to ensure that 
\be \overbrace{{\rm Sym}_k\times {\rm Sym}_k \times\ldots\times {\rm Sym}_k}^p = {\bf 1}\nn\ee
which tells us that \eqn{string} has a unique answer when $R$ is taken to be the maximally symmetric representation. Now our job is to compute it.

\subsection{The Wavefunction as a Correlation Function}

The standard tool to compute correlation functions in WZW models is the Knizhnik-Zamolodchikov (KZ) equation \cite{kz}. Usually this is employed to compute 4-point functions but since we expect a unique solution to \eqn{string}, we can hope to use it in the present case to compute higher-point functions.

\para
The KZ equation reads
\be
\left( \frac{\partial}{\partial z_a} - \frac{1}{k + p} \sum_{b\neq a}^{N} \frac{T^\alpha_a \otimes T^\alpha_b}{z_a - z_b} \right) 
\langle {\cal O}_R(z_1)\ldots {\cal O}_R(z_{N})\rangle = 0
\label{kz}\ee
where $T^\alpha$ is the Hermitian generator for the  $k$-th symmetric representation. These obey the $SU(p)$ algebra 
$[T^\alpha, T^\beta] = i f^{\alpha\beta\gamma} T^\gamma$, with the normalisation $f^{\alpha\gamma\delta}f^{\beta\gamma\delta} = 2p \,
\delta^{\alpha\beta}$, where $p$ appears in its role as the dual Coxeter number of $SU(p)$.

\subsubsection*{Solving the Knizhnik-Zamolodchikov Equation}

We will now show that the KZ equation \eqn{kz} is solved by
\be \langle {\cal O}_R(z_1)\ldots {\cal O}_R(z_{N})\rangle  = \prod_{a<b}^{N} (z_a - z_b)^{-k/p}\ {\cal P}\left[\Phi_{(p)}^k(z,\sigma)\right]\label{corr}\ee
with $\Phi_{(p)}$ is defined in \eqn{phip}. 

\para
It will be useful to first rewrite our ansatz in slightly more concrete form. As discussed up to \eqn{genhalperincoeff}, it's simple to check that, up to an unimportant normalisation, 
\be {\cal P}\left[\Phi_{(p)}^k(z,\sigma)\right] = \Bigg[ {\cal A}&&\!\! \!\!\prod_{1\le a<b \le N/p} (z_a - z_b) \ket{\sigma_{1}}\ldots \ket{\sigma_{1}}\,   \otimes \!\!\prod_{N/p< a<b \le 2N/p} (z_a - z_b) \ket{\sigma_{2}} \ldots \ket{\sigma_{2}} \nn\\ &&\ \ \ \  \otimes \ldots \ldots
 \otimes\!\!\prod_{(p-1)N/p< a<b \le N} (z_a - z_b) \   \   \ket{\sigma_{p}}\ldots \ket{\sigma_{p}}\Big]^k\label{phik}\ee
Here we have placed the first $N/p$ particles in the same spin state, the next $N/p$ in a different spin state and so on. The ${\cal A}$ symbol means that we then antisymmetrise over all particles. 

\para
The  generators $T^\alpha$ in \eqn{kz} can be viewed as acting symmetrically on  what were originally $k$ distinct fundamental factors,
\be
T^\alpha &=& \overbrace{t^\alpha \otimes {\bf 1} \otimes \cdots \otimes {\bf 1}}^k\ +\ \mbox{symmetric permutations}
\nn\ee
with $t^\alpha$ the generator in the fundamental representation. 
The normalisation in the KZ equation ensures that  we have $ t^\alpha_{ij}t^\alpha_{kl} = \delta_{il} \delta_{jk} - \frac{1}{p} \delta_{ij}\delta_{kl}$, where the group indices $i,j,k,l=1,\ldots, p$. 
This means that if the operator $T^\alpha_a \otimes T^\alpha_b$ acts on a state where particles  $a$ and $b$ have the same spin, this tensor product returns the same spin state multiplied by a factor $(1-1/p)$. By contrast, if the two particles have different spins, it returns a superposition of the same state with a factor $-1/p$, and a state with the particles swapped with no factor.

\para
We'll start by considering the action of the non-derivative part of the KZ operator  on ${\cal P}[\Phi_{(p)}^k]$. Expanding out \eqn{phik} will result in a slew of terms, each of them containing $k$ factors. Let's look at one of these terms --- call it $X$. We're going to figure out the coefficient in front of  $X$ after the action of the $T\otimes T$ term in the KZ operator. 
There are two contributions. One arises when $T\otimes T$ acts on $X$ itself. The other arises from $T\otimes T$ hitting other terms in the expansion of ${\cal P}[\Phi_{(p)}^k]$ so that they are mapped into $X$. We deal with these in turn.

\para
 Let's first look at the action of $T\otimes T$ on $X$. Suppose that particle $a$ has the same spin as particle $b$  in $d_b$ of the $k$ factors. Then acting with $T\otimes T$ will, among other things, return $X$ with a coefficient 
\be
-\frac{1}{k+p}  \sum_{b \neq a}^{N} \frac{ k d_b(1-1/p) - k (k-d_b)/p }{z_a - z_b} = -\frac{1}{k+p}  \sum_{b \neq a}^{N} \frac{ kd_b - k^2/p }{z_a - z_b}
\label{a3}\ee
This is our first result.

\para
Now let's see how we can get terms proportional to $X$ by the action of $T\otimes T$ on some other term $Y$. This can occur only if $Y$ differs from $X$  by swapping the spins of $i$ and $j$ in just one factor. Suppose that, within $Y$, particle $i$ has spin $a$ and particle $j$ has spin $b\neq a$. Then the  $T\otimes T$ term in the KZ operator will map $Y$ to $X$ with coefficient
\be 
+\frac{1}{k+p} \frac{ 1 }{z_a - z_b} 
\prod_{(d\neq a|i)} \frac{z_b - z_d}{z_a - z_d}
\prod_{(d\neq b|j)} \frac{z_a - z_d}{z_b - z_d}
\nn\ee
where the notation $(d\neq a|i)$ means that we take the product over all particles $d\neq a$ which carry spin $i$. 
Something nice now happens when this expression is summed over all particles  $j$ which carry  spin $b$; the terms combine to give
\be
+\frac{1}{k+p}\left[ 
-\sum_{(d\neq a|i)} \frac{1}{z_a - z_d}
+\sum_{(d|i)} \frac{1}{z_a - z_d}
 \right] 
\label{a4}\ee
This is our second result. 

\para
The total coefficient multiplying the term $X$  after the action of $T\otimes T$ is then given by the sum of \eqn{a3} and \eqn{a4}. It is
\be
\sum_{b\neq a } \frac{  k/p - d_b}{z_a - z_b} \label{neglect}
\nn\ee
The key point is that this coefficient is precisely cancelled by the derivative term in the KZ equation, since $(z_a-z_b)$ appears with the power  $-k/p + d_b$ in the correlation function \eqn{corr}. Note that the actual coefficient of a given term $X$ typically includes, in addition to \eqn{neglect}, a symmetry factor to account for the fact that $X$ may appear many times in the original expansion of ${\cal P}[\Phi_{(p)}^k]$. This is not relevant for our final result; the same symmetry factor appears every time $X$ arises. This concludes our proof that the correlation function \eqn{corr} indeed solves the KZ equation \eqn{kz}.

\subsubsection*{The Full Wavefunction}

The solution to the KZ equation ensures  that our wavefunction \eqn{final} can be written as the product 
\be \psi_{n+k,n}(z,\sigma) = \prod_{a<b}^{N} (z_a-z_b)^{n+k/p}\,\langle {\cal O}_R(z_1)\ldots {\cal O}_R(z_{N})\rangle\, e^{-\sum_a|z_a|^2/4l_B^2} \label{wf}\ee
where, for once, we've restored the exponential factor common to all lowest Landau level wavefunctions. 
The first factor is, of course, a Laughlin wavefunction and can be expressed as a correlation function for a free compact boson. The fractional exponent is unusual, but has been seen before in constructing Halperin wavefunctions from a CFT \cite{mr,bfradkin} where the same  factorisation into ``spin" degrees of freedom, captured by $\langle {\cal O}_R(z_1)\ldots {\cal O}_R(z_{N})\rangle$, and ``charge" degrees of freedom captured by the Laughlin wavefunction also occurs. 

\para
The full wavefunction can be written as a correlation function in the WZW model with algebra
\be U(p)_{(k+np)p,k} = \frac{U(1)_{(k+np)p} \times SU(p)_k}{{\bf Z}_p}\label{uwzw}\ee
Note that our matrix model describes the Blok-Wen states with $n=1$. Happily, in that case, the WZW model \eqn{uwzw}  indeed arises as the description of the boundary dynamics of  the Chern-Simons  theory with gauge group \eqn{upgauge}.

\para
The $U(1)$ part is described by a compact chiral boson $\phi$ and the correlation function
\be\langle \prod_{a=1}^N e^{i\sqrt{(k+np)/p}\phi(z_i)}\, e^{- \int d^2z' \sqrt{(k+np)/p}\,\phi(z')/2\pi l_B^2}\rangle =  \prod_{a<b}^{N} (z_a-z_b)^{n+k/p}\, e^{-\sum_a|z_a|^2/4l_B^2} \nn\ee
gives the Laughlin part of the wavefunction \eqn{wf} in the usual manner \cite{mr}.

\subsubsection*{Quasi-Holes as Non-Abelian Anyons}

The presence of the $SU(p)_k$ factor ensures that our quantum Hall states have non-Abelian anyons for $k>1$. These quasi-holes are associated to primary operators in the WZW conformal field theory and their properties are expected to be  determined by the fusion rules and braiding inherited from the CFT. Although this story is well known (see, for example, \cite{nayakreview}), we pause here to  point out a few of the more prominent examples. 

\para
\underline{An Example: Ising Anyons}

\para
Ising anyons are well known to appear in the Moore-Read state which is associated to the $SU(2)_2$ WZW model \cite{fradwil}. The primary operators carry spin $s= 0, 1/2$ and $1$. We denote these representations using their dimension ${\bf d} = 2s+1$. As we saw above, we identify the electron with the spin 1, or ${\bf d}={\bf 3}$ representation. The ${\bf d}={\bf 2}$ primary is then identified with the quasi-hole, with the corresponding fusion rules given by
\be {\bf 2}\star{\bf 2} = {\bf 1}\oplus {\bf 3}\ \ \ ,\ \ \ {\bf 2}\star {\bf 3} = {\bf 2}\ \ \ ,\ \ \ {\bf 3}\star{\bf 3} = {\bf 1}\nn\ee
These are the fusion rules for Ising anyons.

\para
\underline{Another Example: Fibonacci Anyons}

\para
Fibonacci anyons are known to arise as the quasi-holes in the ${\bf Z}_3$ parafermionic Read-Rezayi state. This is governed by the coset model $SU(2)_3/U(1)$. As we have seen above, these states arise in our construction as the $SU(3)_2$ WZW model. This is related to the parafermion CFT through level-rank duality and a quotient. The anyon is associated to the primary operator which transforms in the adjoint representation ${\bf 8}$ of $SU(3)$. It is simple to compute the fusion rules in $SU(3)_2$ to find
\be {\bf 8}\star {\bf 8} = {\bf 1}\oplus {\bf 8}\nn\ee
This is indeed the fusion rule for Fibonacci anyons. A nice  review of these objects can be found in \cite{fibreview}.

\para
It remains an open problem to identify these anyonic states directly within the matrix model. For the Laughlin states, it was shown in \cite{us} that the matrix model provides a construction of quasi-hole states which are analytically more tractable than the traditional approach. In particular, the Berry phase computation of \cite{arovas} can be performed exactly, without resorting to the plasma analogy. It seems plausible that the matrix model may also prove useful in understanding the properties of non-Abelian anyons. We hope to return to this in the future.

\section*{Acknowledgements}

Our thanks to Gautam Mandal, Shiraz Minwalla and T. Senthil for useful discussions. DT and CT are grateful to the Tata Institute of Fundamental Research for their kind hospitality while this paper was written. We are supported by STFC and by the European Research Council under the European Union's Seventh Framework Programme (FP7/2007-2013), ERC grant agreement STG 279943, ``Strongly Coupled Systems".

\end{document}